# Heat transfer in MHD flow due to a linearly stretching sheet with induced magnetic field

Tarek M. A. El-Mistikawy

Dept. Eng. Math. & Phys., Faculty of Engineering, Cairo University, Giza 12211, Egypt

**Abstract**

The full MHD problem of the flow and heat transfer due to a linearly stretching sheet in the presence of a transverse magnetic field is put in a self-similar form. Traditionally ignored physical processes such as induced magnetic field, viscous dissipation, Joule heating, and work shear are included and their importance is established. Cases of prescribed surface temperature, prescribed heat flux, surface feed (injection or suction), velocity slip, and thermal slip are also considered. The problem is shown to admit self similarity. Sample numerical solutions are obtained for chosen combinations of the flow parameters.

Keywords: MHD flow; Heat transfer; linearly stretching sheet; induced magnetic field; self similarity.

## 1      Introduction

The problem of the two dimensional flow due to a linearly stretching sheet, first formulated by Crane [1], has a simple exact similarity solution. This invited several researchers to add to it new features allowing for self similarity. As a boundary layer problem, Pavlov [2] added uniform transverse magnetic field. Gupta and Gupta [3] added surface feed (suction or injection). These problems were recognized as being exact solutions of corresponding Navier-Stokes problems by Crane[1], Andersson [4] and Wang [5], respectively. To the Navier-Stokes problem, Wang [6] and Andersson [7], independently, added velocity slip. Fang et al. [8] combined the effects of transverse magnetic field, surface feed and velocity slip.

Heat transfer was treated in several publications, mostly neglecting viscous dissipation and Joule heating (in MHD problems). This allowed self similar formulation in cases of the surface having constant temperature [3,9] or temperature or heat flux proportional to a power of the stretch-wise coordinate $x$ [10,11,12]. Prasad et al. [13] retained viscous dissipation and Joule heating, in case of the surface temperature being proportional to $x^2$.

The abovementioned MHD problems adopted the small magnetic Reynolds number assumption, thus neglecting the induced magnetic field. In [14], it was shown that the full MHD problem, i.e., Navier-Stokes and Maxwell's equations with adherence conditions and appropriate magnetic conditions, allowed for self similarity.

In this article, we extend the work of [14] to the heat transfer problem including viscous dissipation and Joule heating, in cases of quadratic surface temperature or heat flux. Surface feed, velocity and thermal slip, and shear work are also included.

## 2 Mathematical model

An electrically conducting, incompressible, Newtonian fluid is driven by a non-conducting porous sheet, which is stretching linearly in the $x$-direction. At the surface, we consider cases of prescribed temperature or heat flux, and allow for velocity and thermal slip. In the farfield, the fluid is essentially quiescent under pressure $p_\infty$ and temperature $T_\infty$, and is permeated by a stationary magnetic field of uniform strength $B$ in the transverse $y$-direction.

The equations governing this steady two-dimensional MHD flow are

$$u_x + v_y = 0 \tag{1}$$

$$\rho(uu_x + vu_y) + p_x = \rho\nu(u_{xx} + u_{yy}) + \sigma[(B+s)rv - (B+s)^2 u] \tag{2}$$

$$\rho(uv_x + vv_y) + p_y = \rho\nu(v_{xx} + v_{yy}) + \sigma[(B+s)ru - r^2 v] \tag{3}$$

$$s_x - r_y = \sigma\mu[(B+s)u - rv] \tag{4}$$

$$r_x + s_y = 0 \tag{5}$$

$$\rho c(uT_x + vT_y) = k(T_{xx} + T_{yy}) + \rho\nu[2(u_x^2 + v_y^2) + (u_y + v_x)^2] + \sigma[(B+s)u - rv]^2 \tag{6}$$

with the surface conditions

$$y = 0: \ u = \omega x + \lambda_w u_y, \ v = v_w, \ T = T_w + \gamma_w T_y \text{ or } q_w = -kT_y - \rho\nu\lambda_w u_y^2 \tag{7}$$

and the farfield conditions

$$y \sim \infty: \ u \sim 0, \ p \sim p_\infty, \ r \sim 0, \ s \sim 0, \ T \sim T_\infty \tag{8}$$

$(u, v)$ are the velocity components in the $(x, y)$ directions, respectively, and $(r, s)$ are the corresponding induced magnetic field components. $p$ is the pressure and $T$ is the temperature. Constants are the fluid density $\rho$, kinematic viscosity $\nu$, electric conductivity $\sigma$, magnetic permeability $\mu$, specific heat $c$, and thermal conductivity $k$. The stretching rate $\omega$ and the velocity and thermal slip coefficients $\lambda_w$ and $\gamma_w$ are also constant. In the condition for $q_w$, the last term represents the shear work [15]. In the farfield, the condition for $r$ translates the physical requirement of the absence of any current density, while that on $s$ indicates that $B$ stands for the farfield total magnetic field imposed and induced [14].

The problem admits the similarity transformations

$$y = \sqrt{v/\omega}\,\eta,\quad v = -\sqrt{v\omega}\,f(\eta),\quad u = \omega x f'$$

$$s = B\sigma\mu v g(\eta),\quad r = -B\sigma\mu\sqrt{v\omega}\,x g'$$

$$p = p_\infty - \rho\omega v[f' + \tfrac{1}{2}f^2 - \tfrac{1}{2}f^2(\infty)] - \tfrac{1}{2}B^2\sigma^2\mu v\omega x^2 g'^2 \tag{9}$$

$$T = T_\infty + \frac{v\omega}{c}[\theta_0 + \sqrt{\frac{\omega}{v}}\,x\theta_1 + \frac{\omega}{v}x^2\theta_2]$$

where primes denote differentiation with respect to $\eta$. Note that the temperature is quadratic in $x$.

The problem becomes

$$f''' + ff'' - f'^2 - \beta f' = -P_m\beta[g'^2 + (1+P_m g)fg' - (2+P_m g)f'g] \tag{10}$$

$$g'' = f' + P_m(gf' - fg') \tag{11}$$

$$\mathrm{Pr}^{-1}\theta_2'' + f\theta_2' - 2f'\theta_2 = -\beta[f' + P_m(gf' - g'f)]^2 - f''^2 \tag{12}$$

$$\mathrm{Pr}^{-1}\theta_1'' + f\theta_1' - f'\theta_1 = 0 \tag{13}$$

$$\mathrm{Pr}^{-1}\theta_0'' + f\theta_0' = -\frac{1}{\mathrm{Pr}}2\theta_2 - 4f'^2 \tag{14}$$

where $P_m = \sigma\mu v$ is the magnetic Prandtl number, $\beta = \sigma B^2/\rho\omega$ is the magnetic interaction number, $\mathrm{Pr} = \rho v c/k$ is the Prandtl number.

Consistent with the similarity transformations we take the surface values to be

$$v_w = -\sqrt{v\omega}\,f_w \tag{15}$$

$$T_w = T_\infty + \frac{v\omega}{c}[\Theta_0 + \sqrt{\frac{\omega}{v}}\,\Theta_1 x + \frac{\omega}{v}\Theta_2 x^2] \tag{16}$$

$$q_w = \frac{k\omega\sqrt{v\omega}}{c}[Q_0 + \sqrt{\frac{\omega}{v}}Q_1 x + \frac{\omega}{v}Q_2 x^2] \tag{17}$$

where $f_w$, $\Theta_0$, $\Theta_1$, $\Theta_2$, $Q_0$, $Q_1$, and $Q_2$ are prescribed values.

With $\lambda = \lambda_w \sqrt{\omega/v}$ and $\gamma = \gamma_w \sqrt{\omega/v}$, we get the following conditions on the flow variables

$$f(0) = f_w, \quad f'(0) = 1 + \lambda f''(0), \quad f'(\infty) = 0 \tag{18}$$

$$g(\infty) = 0, \quad g'(\infty) = 0 \tag{19}$$

$$\theta_2(0) = \Theta_2 + \gamma\theta_2'(0) \text{ or } \theta_2'(0) = -Q_2 - \Pr\lambda[f''(0)]^2, \quad \theta_2(\infty) = 0 \tag{20}$$

$$\theta_1(0) = \Theta_1 + \gamma\theta_1'(0) \text{ or } \theta_1'(0) = -Q_1, \quad \theta_1(\infty) = 0 \tag{21}$$

$$\theta_0(0) = \Theta_0 + \gamma\theta_0'(0) \text{ or } \theta_0'(0) = -Q_0, \quad \theta_0(\infty) = 0 \tag{22}$$

## 3  Numerical method

We start by solving for $f(\eta)$ and $g(\eta)$, since their problem is uncoupled from the problems for $\theta_2(\eta)$, $\theta_1(\eta)$ and $\theta_0(\eta)$. A closed form solution is not possible, so we seek an iterative numerical solution. In the $n^{th}$ iteration, we solve, for $f_n(\eta)$, Eq. (10) with its right hand side evaluated using the previous iteration solutions $f_{n-1}(\eta)$ and $g_{n-1}(\eta)$, together with conditions (18). Then we solve, for $g_n(\eta)$, Eq. (11) with the known $f_n(\eta)$, together with conditions (19). The iterations continue until the maximum error in $f(\eta_\infty)$, $f''(0)$, $g(0)$ and $g'(0)$ becomes less than $10^{-10}$. For the first iteration, we zero the right hand side of Eq. (10) which corresponds to $g_0(\eta) = 0$.

The numerical solution of the problems for $f_n(\eta)$ and $g_n(\eta)$ utilizes Keller's two point, second order accurate, finite-difference scheme [16]. A uniform step size $\Delta\eta = 0.01$ is used on a finite domain $0 \leq \eta \leq \eta_\infty$. The value of $\eta_\infty = 60$ is chosen sufficiently large in order to insure the asymptotic satisfaction of the farfield conditions. The non-linear terms in the problem for $f_n(\eta)$ are quasi-linearized, and an iterative procedure is implemented; terminating when the maximum error in $f_n(\eta_\infty)$ and $f_n''(0)$ becomes less than $10^{-10}$.

Having determined $f(\eta)$ and $g(\eta)$, we solve the linear problems for $\theta_2(\eta)$, $\theta_1(\eta)$ then $\theta_0(\eta)$, using Keller's scheme on the same grid.

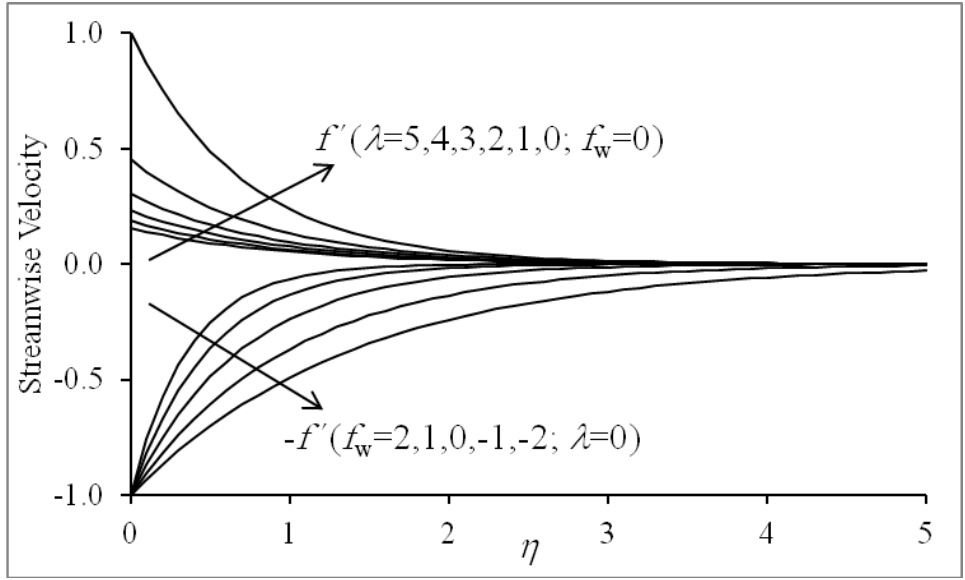

**Fig.1** Streamwise velocity profile

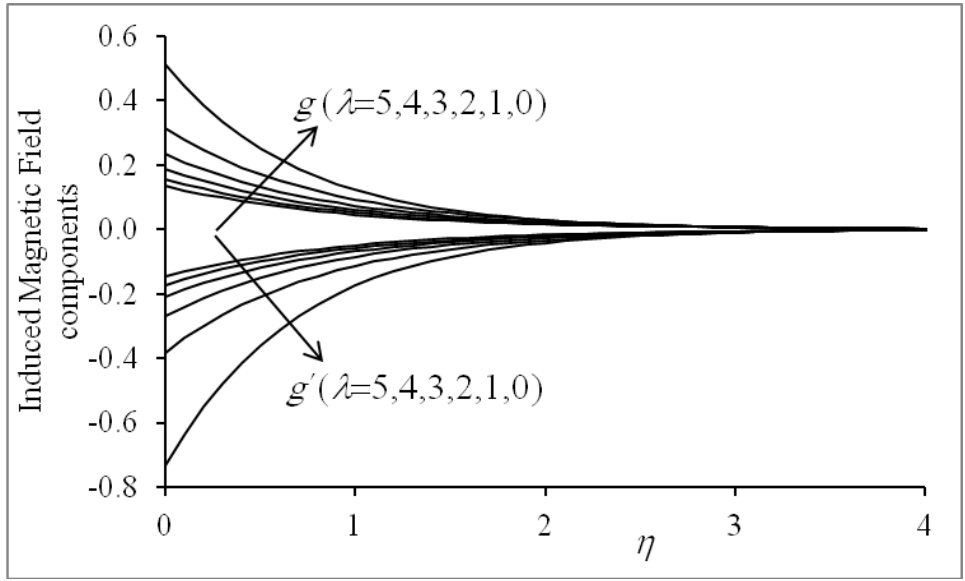

**Fig. 2** Profiles of induced magnetic field components; $f_w = 0$

## 4 Sample results and discussion

The problem for $f(\eta)$ and $g(\eta)$ involves four parameters- $P_m$, $\beta$, $\lambda$ and $f_w$. For $P_m = 0.1$, $\beta = 1$, Fig. 1 depicts $f'(\eta)$ at different values of $\lambda$, when $f_w = 0$, and at different values of $f_w$, when $\lambda = 0$. The corresponding results for $g(\eta)$ together with $g'(\eta)$ are depicted in Figs 2 and 3, respectively. The induced magnetic field is primarily affected by $f'(\eta)$. As $f'(\eta)$ decreases due to higher surface slip or suction rate, both $g(\eta)$ and $-g'(\eta)$ decrease. Tables 1 and 2 give values of the surface shear and the entrainment rate represented respectively by $f''(0)$ and $f(\infty)$, as well as the induced magnetic field components at the surface represented respectively by $g'(0)$ and $g(0)$. Refer to [14] for values of $f''(0)$, $f(\infty)$, $g'(0)$ and $g(0)$ at different values of $P_m$ and $\beta$, when $\lambda = f_w = 0$.

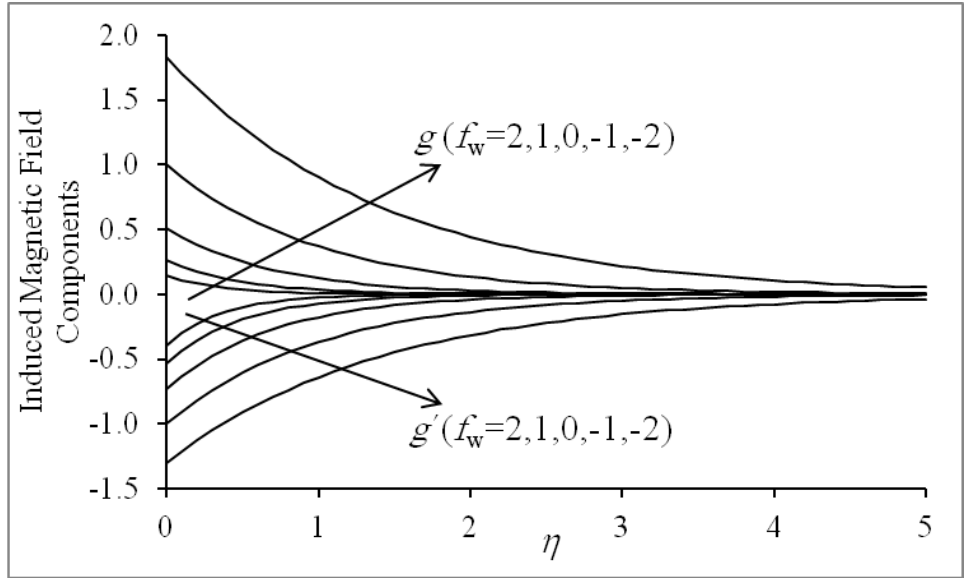

**Fig. 3** Profiles of induced magnetic field components; $\lambda = 0$

**Table 1** Variation of $f''(0)$, $f(\eta_\infty)$, $g(0)$ and $g'(0)$ with $\lambda$; $f_w = 0$

| $\lambda$ | $f''(0)$ | $f(\eta_\infty)$ | $g(0)$ | $g'(0)$ |
| --- | --- | --- | --- | --- |
| 0 | -1.43222 | 0.69822 | 0.51249 | -0.73400 |
| 1 | -0.54904 | 0.37039 | 0.31377 | -0.38201 |
| 2 | -0.34863 | 0.26290 | 0.23363 | -0.26904 |
| 3 | -0.25674 | 0.20564 | 0.18749 | -0.20949 |
| 4 | -0.20357 | 0.16944 | 0.15701 | -0.17210 |
| 5 | -0.16879 | 0.14431 | 0.13524 | -0.14626 |

**Table 2** Variation of $f''(0)$, $f(\eta_\infty)$, $g(0)$ and $g'(0)$ with $f_w$; $\lambda=0$

| $f_w$ | $f''(0)$ | $f(\eta_\infty)$ | $g(0)$ | $g'(0)$ |
|---|---|---|---|---|
| -2 | -0.70972 | -0.59099 | 1.83271 | -1.30070 |
| -1 | -1.00000 | 0.00000 | 1.00000 | -1.00000 |
| 0 | -1.43222 | 0.69822 | 0.51249 | -0.73400 |
| 1 | -2.02629 | 1.49351 | 0.26293 | -0.53278 |
| 2 | -2.75888 | 2.36247 | 0.14369 | -0.39641 |

Results for the temperature constituents $\theta_0(\eta)$, $\theta_1(\eta)$ and $\theta_2(\eta)$ are obtained when $\Pr=0.72$, $P_m=0.1$, $\beta=1$ and $\lambda=f_w=\gamma=0$, in case of prescribed surface temperature $T_w \propto x^a$, and in case of prescribed surface heat flux $q_w \propto x^a$; $a=0$, 1 or 2. Since Eqs. (12)-(14) indicate that $\theta_1$ and $\theta_2$ are independent of the other constituents while $\theta_0$ depends on $\theta_2$, we present, in Figs. 4-9, results for $\theta_0$ at different values of $\Theta_0$ (with $\Theta_1=\Theta_2=0$) or $Q_0$ (with $Q_1=Q_2=0$), for $\theta_1$ at different values of $\Theta_1$ (with $\Theta_0=\Theta_2=0$) or $Q_1$ (with $Q_0=Q_2=0$), and for $\theta_2$ and $\theta_0$ at different values of $\Theta_2$ (with $\Theta_0=\Theta_1=0$) or $Q_2$ (with $Q_0=Q_1=0$), respectively.

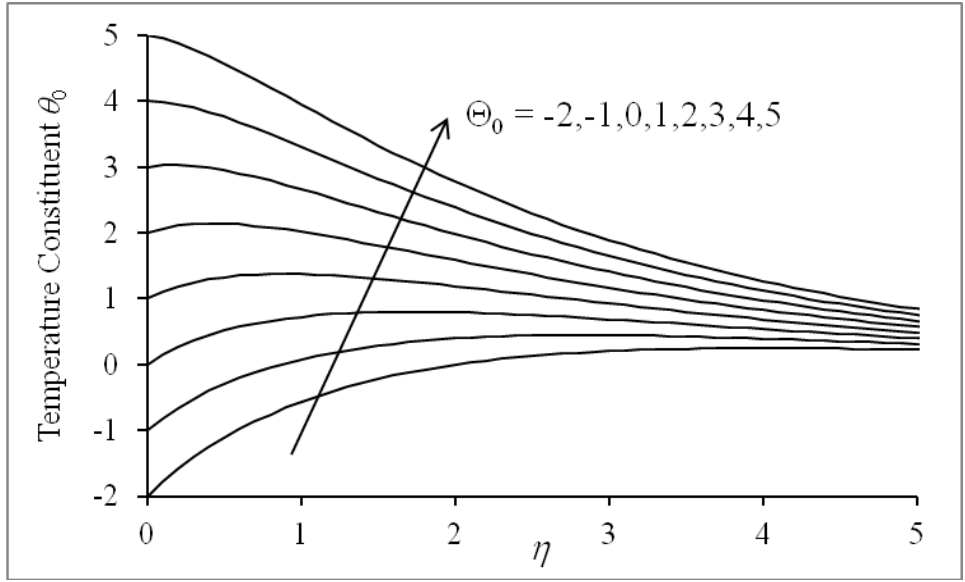

**Fig. 4** Temperature constituent $\theta_0(\eta)$ at different values of $\Theta_0$

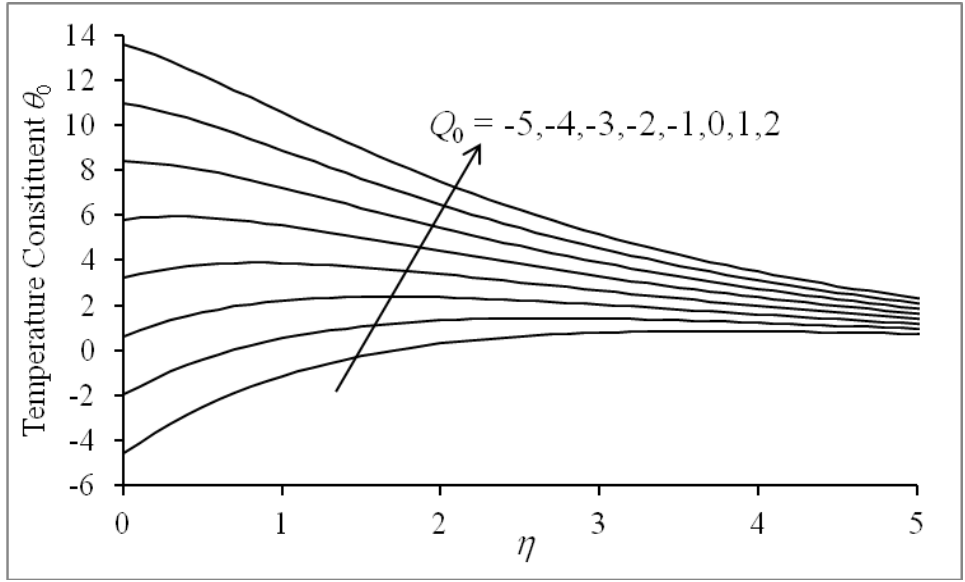

**Fig. 5** Temperature constituent $\theta_0$ ($\eta$) at different values of $Q_0$

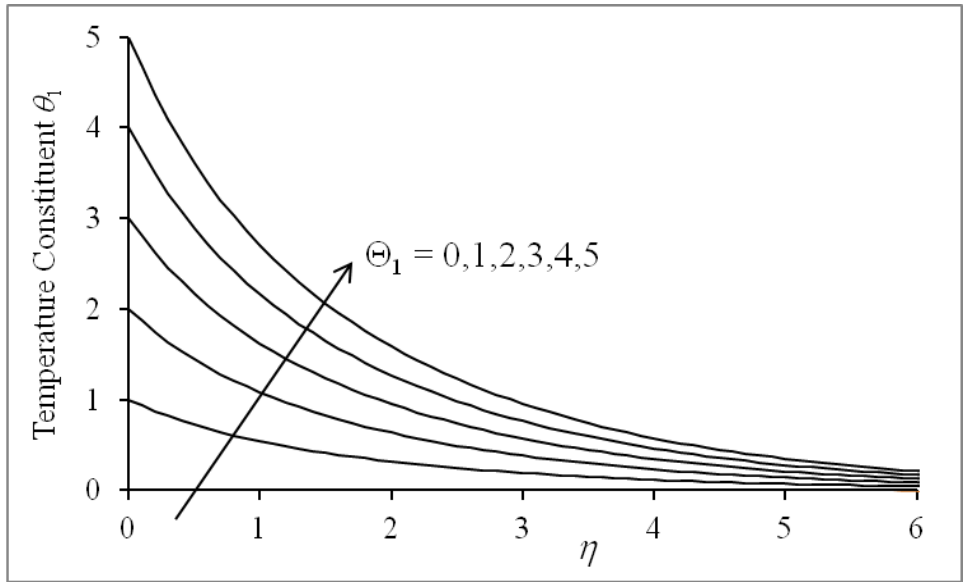

**Fig. 6** Temperature constituent $\theta_1$ ($\eta$) at different values of $\Theta_1$

The following is noticed.

- *Constant surface temperature*. At $\Theta_0 = \Theta_{0c} \approx 4.04660$, $\theta'_0(0) = 0$. For $\Theta_0 > \Theta_{0c}$, $\theta'_0(0)<0$ and $\theta_0$ decreases monotonically with $\eta$, while for $\Theta_0 < \Theta_{0c}$, $\theta'_0(0)>0$ and $\theta_0$ has a peak that gets farther from the surface as $\Theta_0$ decreases.

- *Constant heat flux.* When $Q_0=0$, $\theta_0(0) \approx 8.39634$. As more heat is added to the fluid; i.e. for increasing $Q_0>0$, $\theta_0(0)$ rises and $\theta_0$ decreases monotonically with $\eta$. Removing more heat from the fluid; i.e. for decreasing $Q_0<0$, $\theta_0(0)$ decreases and $\theta_0$ has a peak that gets farther from the surface.

- *Linear surface temperature and heat flux.* Noting that $\theta_1(-\Theta_1)=-\theta_1(\Theta_1)$ and $\theta_1(-Q_1)=-\theta_1(Q_1)$, the presented results for non-negative $\Theta_1$ and $Q_1$ indicate that $\theta_1$ decreases monotonically with $\eta$. Higher $\Theta_1$ results in smaller $\theta_1'(0)<0$, while higher $Q_1$ results in higher $\theta_1(0)$.

- *Surface temperature* $\propto x^2$. At $\Theta_2=\Theta_{2c1} \approx 0.65967$, $\theta_2'(0)=0$. For $\Theta_2 > \Theta_{2c1}$, $\theta_2'(0)<0$ and $\theta_2$ decreases monotonically with $\eta$. For $\Theta_{2c1} > \Theta_2 > \Theta_{2c2} \approx -0.33596$, $\theta_2'(0)>0$ and $\theta_2$ has a peak that gets farther from the surface as $\Theta_2$ decreases. For $\Theta_2 < \Theta_{2c2}$, $\theta_2'(0)>0$ and $\theta_2$ rises monotonically with $\eta$. At $\Theta_2=\Theta_{2c3} \approx -0.61370$, $\theta_0'(0)=0$ and $\theta_0$ drops from its zero surface value to a local minimum then rises to its zero farfield value. Similar behavior of $\theta_0$ is observed for decreasing $\Theta_2 < \Theta_{2c3}$, but with decreasing minimum and $\theta_0'(0)<0$. For increasing $\Theta_2 \geq \Theta_{2c34} \approx -0.33807$, $\theta_0'(0)>0$ and $\theta_0$ rises to a higher peak. For $\Theta_{2c3} < \Theta_2 < \Theta_{2c34}$, $\theta_0$ rises to a peak, falls to zero then to a bottom, and rises again to zero.

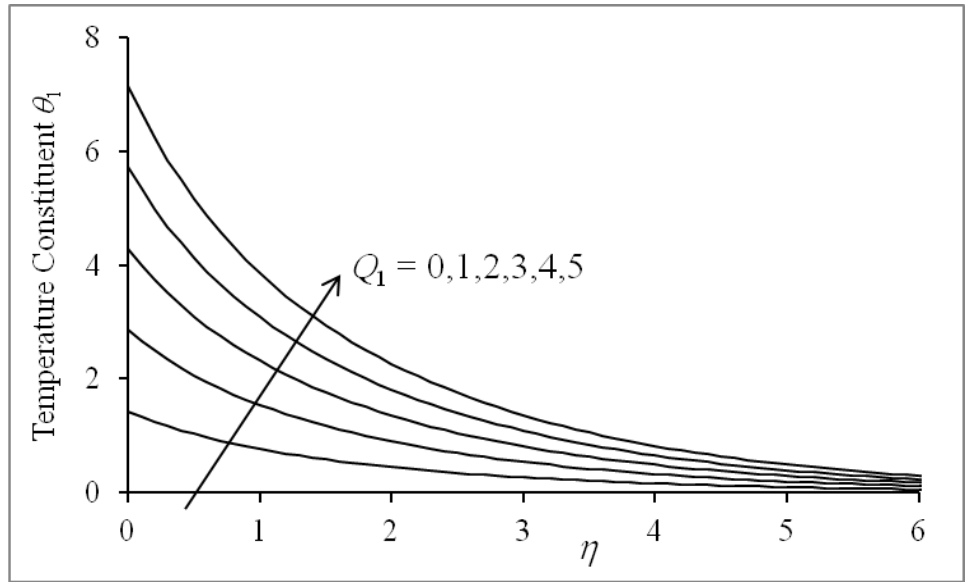

**Fig. 7** Temperature constituent $\theta_1(\eta)$ at different values of $Q_1$

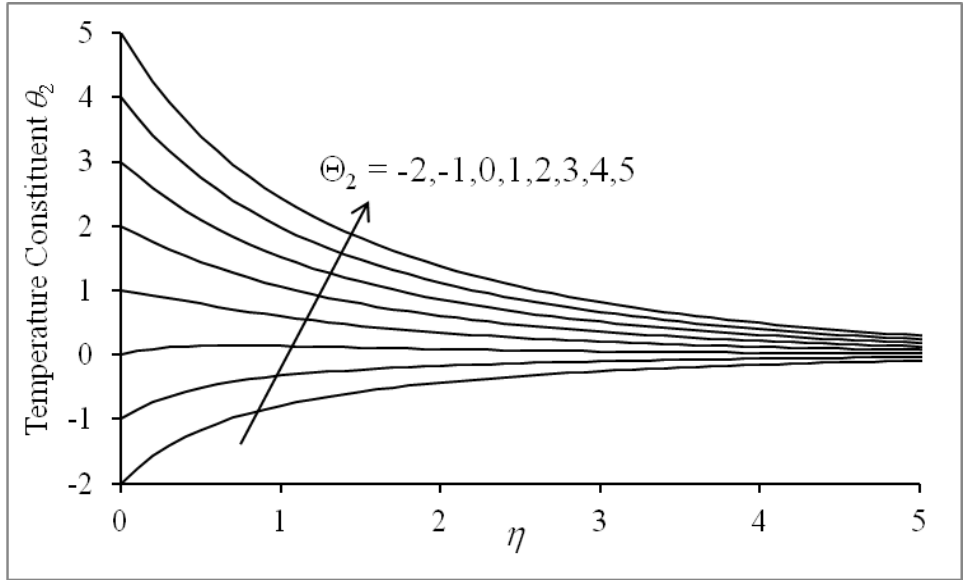

**Fig. 8a** Temperature constituent $\theta_2(\eta)$ at different values of $\Theta_2$

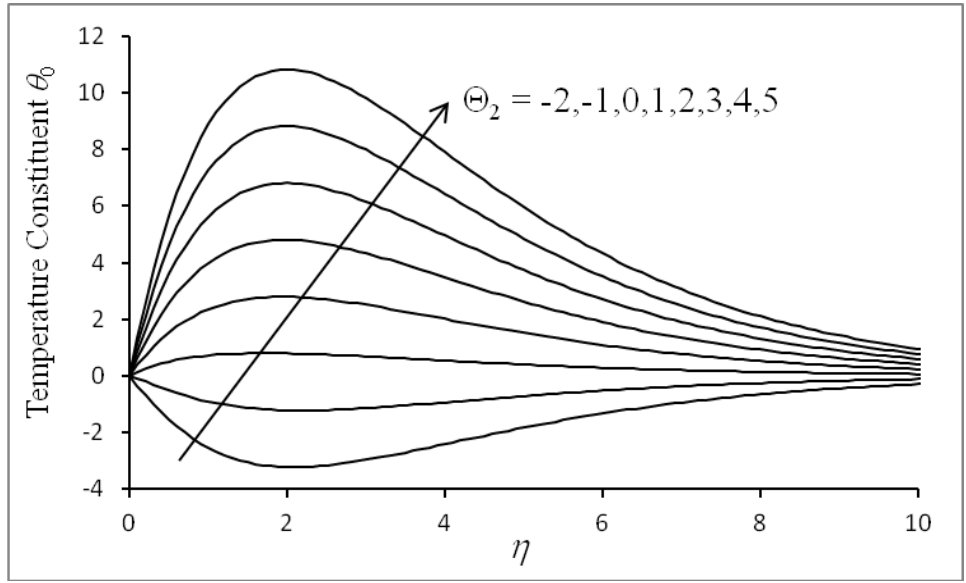

**Fig. 8b** Temperature constituent $\theta_0(\eta)$ at different values of $\Theta_2$

- *Heat flux* $\propto x^2$. When $Q_2=0$, $\theta_2(0) \approx 0.65967$. For increasing $Q_2>0$, $\theta_2(0)$ rises and $\theta_2$ decreases monotonically with $\eta$. For decreasing $0>Q_2>Q_{2c1} \approx -0.96228$, $\theta_2(0)$ decreases from positive to negative values with $\theta_2$ having a positive peak. For $Q_2 \leq Q_{2c1}$, $\theta_2$ rises monotonically from a negative surface value to zero. For $Q_2 \geq Q_{2c2} \approx -0.98114$, $\theta_0$ is monotonically decreasing, while for $Q_2 \leq Q_{2c3} \approx -1.96520$, $\theta_0$ is monotonically increasing from its surface value to zero. For $Q_{2c3} < Q_2 < Q_{2c2}$, $\theta_0$ decreases to a negative local minimum then rises to zero.

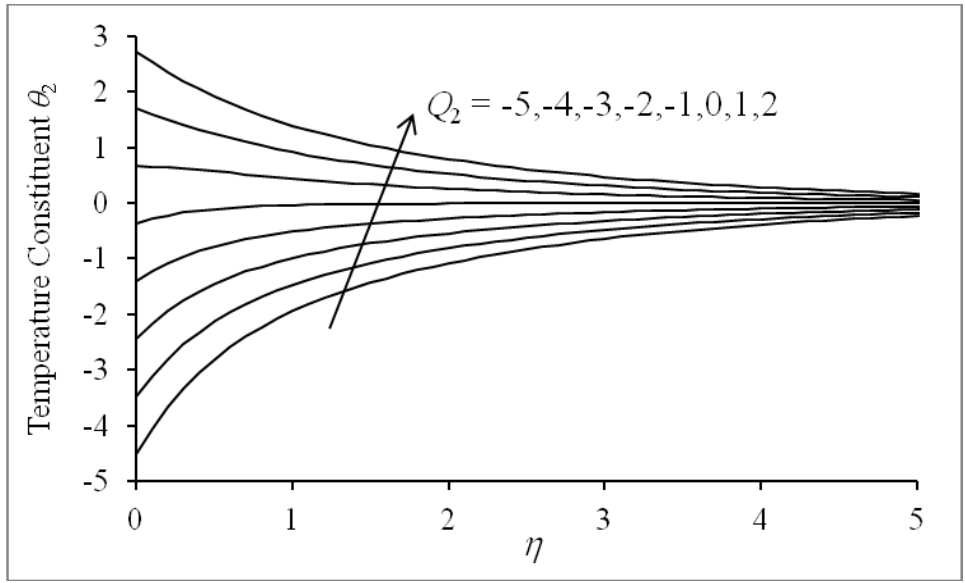

**Fig. 9a** Temperature constituent $\theta_2(\eta)$ at different values of $Q_2$

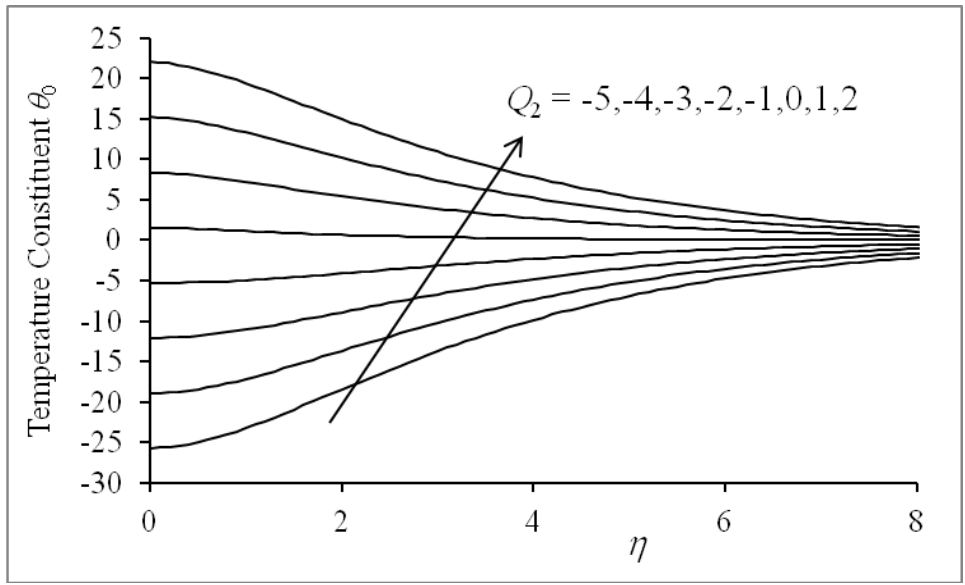

**Fig. 9b** Temperature constituent $\theta_0(\eta)$ at different values of $Q_2$

To complement Figs. 4-9, we give in Table 3 the numerical values of $\theta'_0(0)$ at different values of $\Theta_0$, $\theta'_1(0)$ at different values of $\Theta_1$, and $\theta'_2(0)$ and $\theta'_0(0)$ at different values of $\Theta_2$, and in Table 4 the numerical values of $\theta_0(0)$ at different values of $Q_0$, $\theta_1(0)$ at different values of $Q_1$, and $\theta_2(0)$ and $\theta_0(0)$ at different values of $Q_2$.

**Table 3** Dependence of temperature gradients at surface on surface temperatures; $\gamma = 0$

| $\Theta_0$ | $\theta_0'(0)$ | $\Theta_1$ | $\theta_1'(0)$ | $\Theta_2$ | $\theta_2'(0)$ | $\theta_0'(0)$ |
|---|---|---|---|---|---|---|
| -2 | 2.33284 | -2 | 1.39994 | -2 | 2.57058 | -3.52671 |
| -1 | 1.94703 | -1 | 0.69997 | -1 | 1.60408 | -0.98274 |
| 0 | 1.56122 | 0 | 0.00000 | 0 | 0.63757 | 1.56122 |
| 1 | 1.17541 | 1 | -0.69997 | 1 | -0.32893 | 4.10518 |
| 2 | 0.78960 | 2 | -1.39994 | 2 | -1.29543 | 6.64915 |
| 3 | 0.40379 | 3 | -2.09991 | 3 | -2.26194 | 9.19311 |
| 4 | 0.017978 | 4 | -2.79988 | 4 | -3.22844 | 11.73707 |
| 5 | -0.36783 | 5 | -3.49985 | 5 | -4.19495 | 14.28104 |

**Table 4** Dependence of temperatures at surface on surface heat fluxes; $\lambda = 0$

| $Q_0$ | $\theta_0(0)$ | $Q_1$ | $\theta_1(0)$ | $Q_2$ | $\theta_2(0)$ | $\theta_0(0)$ |
|---|---|---|---|---|---|---|
| 2 | 13.58023 | 2 | 2.85727 | 2 | 2.72898 | 22.04100 |
| 1 | 10.98828 | 1 | 1.42863 | 1 | 1.69433 | 15.21867 |
| 0 | 8.39634 | 0 | 0.00000 | 0 | 0.65967 | 8.39634 |
| -1 | 5.80439 | -1 | -1.42863 | -1 | -0.37499 | 1.57401 |
| -2 | 3.21245 | -2 | -2.85727 | -2 | -1.40964 | -5.24832 |
| -3 | 0.62051 | -3 | -4.28590 | -3 | -2.44430 | -12.07065 |
| -4 | -1.97144 | -4 | -5.71454 | -4 | -3.47896 | -18.89298 |
| -5 | -4.56338 | -5 | -7.14317 | -5 | -4.51361 | -25.71531 |

Table 5 shows the effect of the thermal slip coefficient $\gamma$. The sign of the surface derivative of the temperature constituent determines whether the surface value of the constituent increases or decreases with $\gamma$. Thus, for example, $\theta_2(0)$ increases when $\Theta_2=0$ for which $\theta_2'(0)>0$, and decreases when $\Theta_2=1$ for which $\theta_2'(0)<0$.

**Table 5** Effect of thermal slip coefficient $\gamma$; $\lambda=0$

| $\gamma$ | $\theta_0(0)$ | $\theta_2(0)$ | $\theta_0(0)$ | $\theta_1(0)$ | $\theta_2(0)$ |
|---|---|---|---|---|---|
| 0 | 0.00000 | 0.00000 | 1.00000 | 1.00000 | 1.00000 |
| 1 | 1.72175 | 0.32422 | 2.44335 | 0.58825 | 0.83273 |
| 2 | 3.01106 | 0.43476 | 3.57551 | 0.41668 | 0.77570 |
| 3 | 3.90609 | 0.49050 | 4.36960 | 0.32259 | 0.74695 |
| 4 | 4.55249 | 0.52410 | 4.94569 | 0.26317 | 0.72961 |
| 5 | 5.03861 | 0.54657 | 5.38001 | 0.22223 | 0.71802 |
| Case | $\Theta_0=\Theta_1=\Theta_2=0$ | | $\Theta_0=1$ $\Theta_1=\Theta_2=0$ | $\Theta_1=1$ $\Theta_0=\Theta_2=0$ | $\Theta_2=1$ $\Theta_0=\Theta_1=0$ |

The shear work is represented by the term involving the velocity slip coefficient $\lambda$ in condition (20) for $\theta_2'(0)$. Table 6 demonstrates its importance. Neglecting the shear work reduces the predicted surface temperature.

**Table 6** Effect of shear work; $Q_0=Q_1=Q_2=0$, $\gamma=0$

| $\lambda$ | $\theta_0(0)$ | $\theta_2(0)$ | $\theta_2'(0)$ | $\theta_0(0)$ | $\theta_2(0)$ |
|---|---|---|---|---|---|
| 0 | 8.39634 | 0.65967 | 0.00000 | 8.39634 | 0.65967 |
| 0.2 | 9.69402 | 0.66777 | -0.16337 | 7.84848 | 0.47205 |
| 0.4 | 11.16024 | 0.65306 | -0.21129 | 7.62154 | 0.36953 |
| 0.6 | 12.75638 | 0.63416 | -0.22379 | 7.54814 | 0.30418 |
| 0.8 | 14.46870 | 0.61552 | -0.22308 | 7.56234 | 0.25866 |
| 1 | 16.29111 | 0.59830 | -0.21704 | 7.63203 | 0.22505 |
| Case | Retaining shear work | | | Neglecting shear work | |

On the right-hand-sides of Eqs. (12) and (14), the first terms represent Joule heating and streamwise heat diffusion, respectively, while the second terms represent heat dissipation. Table 7 demonstrates the effect of neglecting these three processes. The predicted heat flux to the surface, represented by $\theta'_0(0)$ and $\theta'_2(0)$, is reduced considerably by neglecting viscous dissipation, less by neglecting streamwise diffusion and lesser by neglecting Joule heating. Neglecting one or more may even predict heat flux in the wrong direction.

**Table 7** Neglect of viscous dissipation, streamwise diffusion, or Joule heating; $\Theta_1 = \Theta_2 = 0$

| $\Theta_0$ | $\theta'_0(0)$ | $\theta'_0(0)$ | $\theta'_0(0)$ | $\theta'_0(0)$ |
|---|---|---|---|---|
| 0 | 0.22395 | 0.92158 | 1.33727 | 1.56122 |
| 1 | -0.16186 | 0.53577 | 0.95146 | 1.17541 |
| 2 | -0.54767 | 0.14996 | 0.56564 | 0.78960 |
| 3 | -0.93348 | -0.23585 | 0.17983 | 0.40379 |
| 4 | -1.31929 | -0.62166 | -0.20598 | 0.017978 |
| 5 | -1.70510 | -1.00748 | -0.59179 | -0.36783 |
| Case | Neglecting Dissipation | Neglecting Diffusion | Neglecting Joule heating | Retaining all |
| $\theta'_2(0)$ | 0.22323 | 0.63757 | 0.41434 | 0.63757 |

## 5  Conclusion

The problem of the flow due to a linearly stretching sheet in the presence of a transverse magnetic field has been formulated to include surface feed, velocity and thermal slip. The problem has been shown to admit self similarity of the full MHD fluid flow equations. Included in the thermal equation and conditions are physical processes such as viscous dissipation, Joule heating, streamwise heat diffusion, and shear work which were traditionally ignored or approximated. The importance of these processes as well as the significance of the formulation has been established through samples of numerical results.

## References

[1] L.J. Crane, Flow past a stretching plate, J. Appl. Math. Phys. ZAMP 21 (1970) 645–647.